# A Systemic and Cybernetic Perspective on Causality, Big Data, and Social Networks in Tourism

*Miguel Lloret-Climent, Andrés Montoyo-Guijarro, Yoan Gutierrez-Vázquez, Rafael Muñoz-Guillena, Kristian Alonso-Stenberg*

**Abstract**
**Purpose** The aim of this paper is to propose a mathematical model to determine invariant sets, set covering, orbits and, in particular, attractors in the set of tourism variables. Analysis was carried out based on an algorithm and applying an interpretation of chaos theory developed in the context of General Systems Theory and Big Data.
**Design/methodology/approach** Tourism is one of the most digitalized sectors of the economy and social networks are an important source of data for information gathering. However, the high levels of redundant information on the web and the appearance of contradictory opinions and facts produce undesirable effects that must be cross-checked against real data. This article sets out the causal relationships associated with tourist flows in order to enable the formulation of appropriate strategies.
**Findings** Our results can be applied to numerous cases, for example, in the analysis of tourist flows, these findings can be used to determine whether the behaviour of certain groups affects that of other groups, as well as analysing tourist behaviour in terms of the most relevant variables.
**Originality/Value** The technique presented here breaks with the usual treatment of the tourism topics. Unlike statistical analyses that merely provide information on current data, we uses orbit analysis to forecast, if attractors are found, the behaviour of tourist variables in the immediate future.

## 1. INTRODUCTION

Our basis for studying tourism and tourist destination management was the General Systems Theory (GST) of Ludwig von Bertalanffy (1968, 1972) who defined a system as "a set of elements standing in interrelation among themselves and with the environment." This means that tourism can be seen as an interrelated and integrated system. From the known data – after a detailed www search – it seems that the first precedent of applying GST to tourism, was advocated by Cuervo (1967), who comments that population movements generate relations, services and facilities that interact among themselves. Recently Tănasie (2016), brings to light the agreements and conventions that the Socialist Republic of Romania has entered into from 1969 to 1972 with other countries from the communist bloc. Romania's entry into the tourist circuit implied not only a series of investments made by the authorities but also their involvement in attracting partners in the field.

Beni (1999) defined the tourism system (SISTUR) as "a set of procedures, ideas and principles that are ordered logically and related to the intention of seeing how tourism operates as a whole." He attempted to conceptualise, describe and define the tourism system, identifying the system's components, its cause-effect relationships, and the rise of controlling and dependent subsystems.

The World Tourism Organization (UNWTO), Sancho (2008), Bosch and Merli (2013), propose an approach to the process of analysing tourism activity known as the FAS model, i.e., Factors, Attractors and Support Systems. 'Factors' refers to the resources of the tourist destination that form the production structure. Resources may be natural (such as climate, terrestrial); human (cultural heritage, labour conditions, human capital); and capital (such as physical and financial). 'Attractors' are elements that that make up the tourist attractions of the destination. The attractors are divided into natural (sun and beach, natural spaces), cultural (events, historical heritage) and artificial (shopping, entertainment, business and conventions). 'Support systems' complete the FAS model, among which are included transportation (land, air), hospitality (accommodation, restaurants, destination organizations, complementary services (health, safety, etc.)

Tourism can be analysed as a dynamic system. Forrester's (1971) systems dynamics aim to capture the largest number of feedback loops. System perturbation —by integrating or rejecting variables— enables observation of the model's evolution or resistance to change, as well as predicting the effect of the interaction of the changing variables. A problem to consider, is that all the variables are interrelated, however, inoperative and unnecessary variables must be avoided to prevent system fails. Other approaches to using systems dynamics for studying tourism phenomena are found at OECD (1977), Hamal (1997), Serra (2007), Friedel and Chewings (2008), Vázquez-Ramírez et al. (2013). For example, the dynamic model of tourism, simulates and forecasts the behaviour of tourism in the year 2020.

Presently, tourism phenomena cannot be accurately analysed without considering the impact of the internet. The Internet currently has more than 3.8 billion users (http://www.internetworldstats.com/stats.htm, June 2017), which means that 49.7% of the world's population is connected to the World Wide Web and is, therefore, consuming and generating online information. Between 2000 and 2017, the growth of online users has been exponential—976% between 2000 and 2017. The applications provided by the internet range from online games and downloads of computer apps to contexts in which the users themselves are actively involved in the creation of content. The latter case involves the Web 2.0 (or the social Web). The Social Web has provides opportunities for the creation of new websites where users play a key role participating, interacting and exchanging information with other users in, for example, blogs, social networks, microblogs.

Given the attractiveness of Web 2.0, there has been a significant increase in the amount of User Generated Content (UGC). For example, by consulting the available data of two of the most well-known social networks, Twitter and Facebook, it can be verified that Twitter has more than 310 million active users, who generate 500 million tweets a day (http://www.internetlivestats.com/twitter-statistics, December 2017). Facebook has more than 1.5 billion users (http://www.trecebits.com/2016/01/28/facebook-ya-tiene-1-590-millones-de-usuarios/, January 2016) and generates more than 78 million pages (http://www.statisticbrain.com/facebook-statistics, December 2017). In addition to the pages generated by Twitter and Facebook, if we include other websites, other types of social networks, Web pages, encyclopedias, blogs, forums, multimedia content, there are more than 4.5 billion indexed web pages (http://www.worldwidewebsize.com/ December, 2017).

This work is a case study based on Spanish tourism sector. The Spanish National and Integrated Tourism Plan *(Plan Nacional Integral de Turismo -PNIT* 2012-2015 approved by SEGITTUR 2015, stressed the importance of giving priority to mature destinations: "Numerous destinations, protagonists of the historical growth of our tourism, face a systemic problem. Certain destinations have engaged in a vicious circle of increased price competition with destinations that have lower cost structures. This prevents unsustainable market prices, thereby eroding operating margins, discouraging reinvestment and causing a reduction in perceived quality standards, which in turn reduces the willingness to buy and places greater emphasis on prices.

The tourism sector is characterized as intensive in both the generation and use of information, and increasingly generates digital information that can be analysed using big data terms, which refer to a data set whose size exceeds the ability to search, capture, store, manage, analyse, etc. The technology allows the connection of different physical elements, services and spaces, and in the near future all the territorial elements will be interconnected, facilitating real-time approximation and the generation of a huge amount of data. The present challenge is threefold: i) develop a specialized approach that allows us to understand and search for correlations in a world dominated by data; ii) develop the potential to predict phenomena; iii) find patterns of behaviour, trends and respond to market behaviour proactively. However, the main drawback of this large amount of available information is the complexity of analysing it, especially when it is necessary to interpret information formulated in natural language.

The aforementioned challenges alongside the existence of redundant information, contradictory opinions and facts that may arise, cause several undesirable effects, such as the difficulty of judging brands appropriately. Moreover, users will need to invest more time than what should be necessary to analyse the information and select what is of interest to them. One way to reduce the time needed to analyse large amounts of information is through the use of Human Language Technologies (HLT) (Batrinca and Treleaven, 2015; Chen, Lin and Yuan, 2017). These technologies are fundamental to handle such information efficiently and effectively, since they are capable of processing human language automatically. HLT is a subdiscipline of Artificial Intelligence, which formulates computationally effective mechanisms to facilitate human-machine interrelation, and enables a more fluid and less rigid communication than formal languages. The tools and resources developed in recent years have made it possible to improve the search, recovery and extraction of information, classification of texts, detection and mining of opinions (Gutierrez et al 2015, Abreu et al., 2017, Montoyo, Martínez-Barco and Balahur, 2012). Furthemore, the synthesis of information and the intermediate processes involved in each of these tasks, such as semantic analysis (Gutiérrez, Vázquez and Montoyo, 2016, 2017), which is key to the proper functioning of HLT, have also improved data search, recovery and extraction.

The paper introduces an original approach to the tourism within life support system by proposing concepts that are discussed and defined and that will provide cyberneticians (Wiener, 1948; Vallee, 1995) and systemists with a revised view of systemic thinking. We studied fourteen types of accommodation (according with the Normative of the General Secretariat of Tourism in Spain): Five-gold-star hotels, four-gold-star hotels, three-gold-star hotels, two-gold-star hotels, one-gold-star hotels, three- and two-silver-star hotels, one-silver-star hotels, luxury and first class campsites, second class campsites, third class campsites, total hotel establishments, total campsites, tourist apartments, rural tourism accommodation establishments.

## 2. BASIC CONCEPTS

Big data techniques come from various fields such as statistics, computer science, applied mathematics and economics, and these techniques have recently been deployed in Spain by INVAT.TUR (2015), in the study "Big data, challenges and opportunities for tourism" where the following techniques (figure 1) were adopted:

i) Association rules. Allow the discovery of relevant relationships or association rules, between data variables. For example, these association rules enable researchers to discover what products are frequently bought together.
ii) Network analysis. Techniques used to characterize the vertices of a network. For example, the analysis of social networks determines connections between individuals in a community, how information travels, or who has the most influence over whom.
iii) Predictive models. Techniques by which a mathematical model, is created or chosen to predict the probability of a result. For example, predicting the entry of foreign tourists and their anticipated overnight stays.
iv) Analysis of time series. Set of statistical and signal processing techniques for the analysis of data sequences at correlative time points for the extraction of patterns and of the significant characteristics of the data. For example, to find consumer behaviour patterns in a given context.

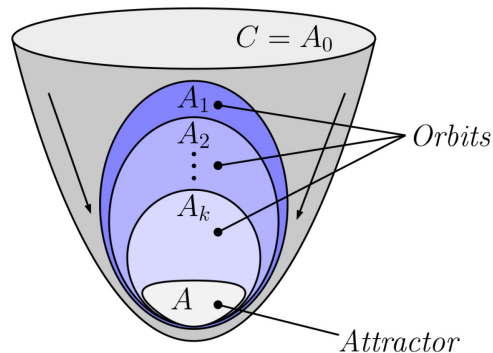

Figure 1. Orbit and attractor (hybrid technique)

This study adopts a hybrid technique that embraces both statistic and social data. IT (information technology) is its most impressive form, one that immediately enters into tourist praxis, keeping the focus on mobility. In parallel, Liebman (2001) examined the concept of mobility as a mobility circle.

It uses both real data obtained from the official website of the Spanish National Institute of Statistics and GPLSI Social Analytics tool (http://socialanalytics.gplsi.es), a platform for visualizing what is being said in real-time about some selected entities in social networks. This hybrid technique creates a dynamic network that discovers relationships between variables. The interaction of these components will explain the behaviour of the data in the structure of the tourism system and may provide future predictions. For example, the carrying capacity of the natural environment in a tourist destination will facilitate decision making that prevents negative consequences for tourism.

In the last few years a small but steadily growing strand of literature has started to consider tourism systems, and particularly a tourist destination, from a "complex systems science" perspective. Many authors have employed complexity and chaos-based approaches to tourism, Edgar and Nisbet (1996), McKercher (1999), Russell and Faulkner (1998), Baggio and Sainaghi (2011). Intervening in the causal series that generate effects and consequences among their variables allows us to obtain the feedback loops, which identify possible attractors in the tourism support system. More specifically, we will focus on the previously mentioned FAS model forward by the UNWTO rather than natural attractors that cannot be controlled or modified. For instance, we cannot influence the main natural attractor of Spain as a destination: sun, climate and beaches. Hence, we search for attractors in the Support Systems. More specifically, we look for tourist attractors of the destination that refer to the place of overnight stay. A stay in a destination can be ruined by poor service at the airport, or because some of the elements that make up the "package" are not well coordinated. In this article, we will focus on the need for accommodation, in which the tourist companies involved are hotels, campsites, apartments and rural tourism establishments, among others.

In chronological order, we proceeded as follows:
1) The data were entered in the SPSS software package to obtain the statistical correlations between the overnight stays of various establishments.
2) We know that a correlation between two units is a necessary but insufficient condition for the existence of a causal relationship between the two variables. To ensure that the relationship between the various types of residences is "causal" rather than "coincidental," we used the concept of the conditioning factor, conditional probabilities and categorical variables were used to define the direction of causality between the different types of accommodation: *A* is a conditioning factor of *B* if $P(B/A) > P(B/\overline{A})$ . This property is used to define the direction of causality between the different types of accommodation.
3) This gave us a directed graph in which the existence of a direction from the *i* factor to the *j* factor means that the *i* factor is a cause of the *j* factor.
4) We applied chaos theory results to this graph to obtain invariant sets, covering between sets, orbits and above all, the attractors in the set.

## 3. IMPLEMENTATION

For a particular individual, we agree with tourism specialists that "Staying at one type does not cause staying at another", but in this manuscript we talk about the overall behaviour of tourists and for example, an improvement in economic conditions or increased age may indicate a tendency of such behaviour.

We have decided to use this specific methodology, which is original, innovative, daring and, from our point of view, valid to interpret the choice of tourist accommodation and the most appropriate one for the kind of research we are conducting. To make the data uniform, we chose monthly data for overnight stays in all these various types of establishments from 2006 to 2017, giving us associated data for one hundred and forty-four months.

These data allowed us to obtain the correlation coefficients between each pair of establishments, studied using the SPSS program.

Pairs of establishments with a correlation coefficient greater than or equal to 0.7 were selected, we have chosen a correlation greater than 0.7 because we consider it a coefficient high enough for there to be an interesting network of causes and effects. If we had chosen a correlation less than 0.7 all types of accommodation would be related to each other, so we could not obtain interesting conclusions. In the same way, if the correlation coefficient is greater than 0.7 the network would be too simple so neither would provide information. Most relationships are simple except for some which were bidirectional.

We used the associated graph in the Fig. 1 below to ascertain the intrinsic details of this system. In the graph, the 30 cause-effect relationships obtained where the origin of the arrow represents the cause and effect are represented end. For example the expression: Rural tourism accommodation → one-gold-star hotels means that people who stay in rural tourism accommodation, "generally", by age or by improving their economic status, just overnight in hotels a gold star.

We analysed the sets associated with different types of establishments, determining the coverings between sets, invariant sets, orbits, loops, Esteve and Lloret (2006 a,b) and above all the attractors, Esteve and Lloret (2007). The concept of orbit determines the zone of influence of the joint behavior of tourists between the different types of accommodation. We used an algorithm developed based on these concepts to facilitate the calculations, Lloret et al (2009).

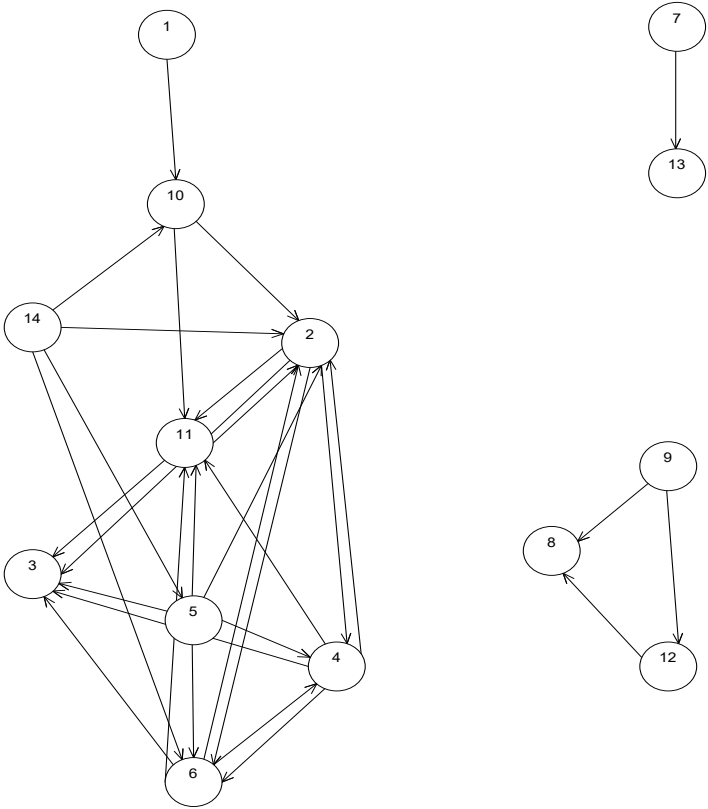

Figure 2. Associated graph for overnight stays in various types of establishments.
Five-gold-star hotels (1) four-gold-star hotels (2), three-gold-star hotels (3), gold-two-star hotels (4), one-gold-star hotels (5), three-and two-silver-star hotels (6), one-silver-star hotels (7), luxury and first class Campsites (8), second class Campsites (9), third class Campsites (10), all hotel establishments (11), Total Campsites (12), tourist apartments (13), rural tourism accommodation establishments (14).

In order to enrich the results obtained, we set up a social monitorisation in Spanish language, (12/12/2017 00:00 to 31/12/2017 23:59), of the same entities listed at Figure 2. This will allow the discovery of demographic correlations provided by social media, and to this end, we employed GPLSI Social Analytics tool. Among the different views and statistics, this tool offers a set of demographic and subjective analytics to estimate how the entities are being valued by the users, using HLT generally focused sentiment analysis (Fernández et al., 2017a). This platform generates real-time reports about some specific entities (people, products, organisations, topics, etc.), summarising what is being said about them and how they are being valued.

## 4. RESULTS

Focused on the correlation mathematic model, we calculated the orbits of the various types of establishments with the following results:

The establishments {Three-gold-star hotels}, {Luxury and first class campsites} and {Tourist apartments} lack orbits but are included in the orbits of other establishments (their rows in the matrix are zero).

The sets {Luxury and first class campsites}, {Second class campsites} and {Total campsites} form a block that is isolated from the other establishments, as shown in the associated graph in Fig. 1. Likewise, the sets {One silver star hotels} and {Tourist apartments} are another block that is separate from the other establishments.

The other types of establishments interact, as can be seen in the associated graph. A more detailed study of the sets shows us the orbits and enables us to obtain the basin of attraction which is formed by the set:

C={Five-gold-star hotels, four-gold-star hotels, three-gold-star hotels, two-gold-star hotels, one-gold-star hotels, three- and two-silver-star hotels, third class campsites, rural tourism accommodation establishments, total hotel establishments} and the attractor is formed by the intersections of the orbits of the sets located in the basin of attraction. The basis of attraction is the set of variables involved with causes and effects in obtaining the attractor set. The attractor set is a loop and an invariant set satisfying the condition that when the effects reach this set not leave itself but rather remain indefinitely in the set, which in our case is the set:

A= {Four-gold-star hotels, three-gold-star hotels, two-gold-star hotels, three- and two-silver-star hotels, total hotel establishments}

In addition, based on GPLSI Social Analytics we obtained a set of analytics related to the same 14$^{th}$ accommodation types. 14.300 messages were collected which came from around 10.800 Twitter users. The potential audience of those messages is around 109.700.000 users, a not insignificant figure.

Table 1 shows the reputation of the different accommodation types. "Hotel establishment", "Rural tourism accommodation establishment" and "Campsites" are global overviews of each category whereas the rest of the accommodation types focus on specific aspects or levels of those global categories. The reputation is measured by a formula described in (Fernandez et al., 2017b), which considers sentiment polarity and other features mentioned, like followers, and others.

Furthermore, the number of entity mentions is counted, which is the same as the number of messages related to an accommodation type. As observed, some accommodation types like "Luxury and first class Campsites" that have a high reputation but have only been mentioned a few times, which is not enough in big data terms. The most mentioned general accommodation types were "Hotel establishment", "Rural tourism accommodation establishment" and "Campsites" respectively.

Finally, when the origin of the active users is observed it reveals that in Spain the "Hotel establishment" is the most popular on social networks, followed by "Rural tourism accommodation establishment" and "Campsites" respectively.

Many conclusions can be drawn from the analysis of the data mined; however, our aim is focused on discovering correlation indicators among accommodation types.
Interesting convergences were found between emotions and sentiment polarities expressed by users.

| Reputation | | Mentions | |
|---|---|---|---|
| Luxury and first class Campsites | 90.1 | Hotel establishments | 13.7K |
| Five-gold-star hotels | 54.5 | Rural to… | 353 |
| Rural tourism accommodati… | 53.2 | Campsit… | 283 |
| Three-gold-star hotels | 43.8 | Five-gol… | 207 |
| Campsites | 38.1 | Tourist … | 10 |
| Tourist apartments | 37.4 | Four-gol… | 6 |
| Hotel establishm… | 7.4 | Three-g… | 1 |
| Gold-two-star hot… | 0 | Luxury … | 1 |
| One-silver-star h… | 0 | Gold-tw… | 0 |
| Three-and two-si… | 0 | One-silv… | 0 |
| Second class Ca… | 0 | Three-a… | 0 |
| Third class Cam… | 0 | Second … | 0 |
| One-gold-star hot… | 0 | Third cl… | 0 |
| Four-gol… | -99.9 | One-gol… | 0 |

Table 1. Global views about accommodation.

For instance, the data revealed negative opinions about hotels. Nevertheless, more in-depth analysis indicated that the negative mentions referred to hotels in general, without any specific classification. Therefore, we carried out a detailed inspection of the hotel related messages and we found a large proportion of messages linked to the Catalonia Crisis (2017) that negatively affected hotel reputation in Spain.

People generally expressed negative sentiments towards hotels; however, at the same time the tone was detected as enthusiastic and therefore classified under the "Happiness category". For example, we listed, at least 127 messages expressing negative sentiments in which, some users also expressed their enthusiasm in the same message.

Apart from the data previously mentioned, we explored in detail the data relating to "Five gold star hotels" category. This type of accommodation based on the social analysis done by GPLSI Social Analytics, deals with 188 positive versus 18 negative mentions. This is the reason why their reputation reaches + 54.5 points, see Table 1. The negative reputation of "Four gold star hotels" is not relevant given that only one user message was spotted resulting in five retweets. This main user message belonged to a person who was expressing his/her bad mood while in a hotel, but it was not related to any hotel in particular. This represents a difficult challenge that needs to be resolved by NLP technologies due to the lack of data related to "Four gold star hotels". Thus, our recommendation for analysing subjective data is to consider only large quantities of data so that noisy information can be given minimal priority. The rest of the accommodation types obtained more positive than negative comments.

## 5. INTERPRETATIONS AND CONCLUSIONS

The technique we have set out in this paper is a combination of statistics and chaos theory as, on the one hand, we used correlation coefficients and calculation of probabilities to prepare the data for study, obtaining pairs of cause-effect relationships with their associated graph, and on the other we applied results in chaos theory to the cause-effect relationships obtained.

The most important aspect of our procedure was that, unlike statistical methods that provide information about data at the present time, with the orbital analysis if we find attractors we can anticipate how clients of the various types of establishments behave in the immediate future as the causes chronologically precede their effects.

In our case, the invariant set {luxury and first class campsites, second class campsites, total campsites} behaves as an endogamous set in which the elements only relate to each other and maintain their structure and status over time. Interestingly, third class campsites do not interact with any other type of campsite. However, they do interact with the various types of hotel establishments, which in principle is surprising.

The attractor set A= {Four-gold-star hotels, three-gold-star hotels, two-gold-star hotels, three- and two-silver-star hotels, total hotel establishments} represents the goal pursued by tourists staying overnight in the sets located in the basin of attraction.

In this case, four interesting results should be noted:
The first, mentioned above, is the fact that tourists staying overnight in third class campsites end up becoming candidate tourists for establishments located in the attractor set.

The second interesting result is that something similar occurs with tourists who stay in rural tourism accommodation establishments, who are also potential clients for establishments located in the attractor.

The third notable result lies with the hotels in the extreme categories: Five-gold-star hotels and one-gold-star hotels, which even if they are in the basin of attraction are not located in the attractor. It is as if many of the guests at five-gold-star hotels eventually prefer staying in establishments with four stars or less, and as if guests at one-gold-star hotels eventually prefer hotels with two or more gold stars, which implies a more homogeneous behaviour among the clients. This may also be due to the hotels' quality policies.

Finally, the results indicate that the guests at one-silver-star hotels are potential customers for tourist apartments.

These results are useful for the marketing policies of various types of institutions - national, regional and local - involved in tourism promotion. Therefore, the resources available could be optimised considering the medium and long-term objectives, rather than the short-term goal.

In conclusion, from our point of view it would be preferable to establish advertising campaigns about the sets located in the basin of attraction which are not part of the attractor (Five-gold-star hotels, one-gold-star hotels, third class campsites, rural tourism accommodation establishments), since in this way the establishments located in the attractor set would also benefit.

Tourism is a mainstay in the evolution of the Spanish economy and society, and needs to renew its growth criteria to ensure and maximise its contribution to social wellbeing, shaping the new era of tourism, defined by technological change, environmental responsibility and the demands of the new society.

The technique presented here breaks with the usual treatment of the tourism topics and its methodology consists in: our study we need to reduce their complexity in order for the destinations to be manageable and to be able to describe their interdependencies, based on the data published by the Spanish National Statistics Institute (http://www.ine.es/) for overnight stays in various types of hotels, various categories of campsites, apartments and rural tourism establishments.

The social analysis has delivered some key conclusions. First, the most popular accommodation category in this study is hotel, although non-positive expectations shade its reputation. As expected, five-gold-star hotels attained the highest standing. Second, after analysing demographic correlations we found the more active Spanish autonomous regions in social media regarding tourism are both: C. de Madrid - Castilla La Mancha and Catalunya. This means that both regions are very suitable for disseminating social media promotional campaigns. However, there are other Spanish regions that need to promote their respective accommodation sectors, at least in social media terms, such as Asturias-Cantabria, Castilla y Leon (see Table 1). Finally, a noteworthy revelation was the negative impact of political events – i.e., the Catalonia Crisis, which negatively affected Spanish hotel reputation in social networks. Clients perceived an uncomfortable city environment due to the extensive popular protests in close proximity to their hotels.

**ACKNOWLEDGMENTS**

This research has been is partially funded by the Office of the Vice President of Research and Knowledge Transfer, University of Alicante, supported this paper under project (GRE15-13).

This research has been is partially funded by the University of Alicante, Generalitat Valenciana, Spanish Government ("Ministerio de Economía y Competitividad") through the projects REDES (TIN2015-65136- C2-2-R) and "Plataforma inteligente para recuperación, análisis y representación de la información generada por usuarios en Internet" (GRE16-01).